\documentclass{article}
\usepackage{spconf,amsmath,graphicx,hyperref}

\usepackage{algorithmic}
\usepackage{graphicx}
\usepackage{textcomp}
\usepackage{xcolor}
\usepackage{hyperref}
\usepackage{orcidlink}
\usepackage{booktabs}
\usepackage{subcaption} % Add this in your preamble
\usepackage{bm}

\definecolor{MS-color}{RGB}{128,0,128}

\renewenvironment{thebibliography}[1]{%
   \begin{oldthebibliography}{#1}%
     \setlength{\itemsep}{0.5pt}%
}%
{%
   \end{oldthebibliography}%
}

\title{DISCRIMINATING REAL AND SYNTHETIC SUPER-RESOLVED AUDIO SAMPLES USING EMBEDDING-BASED CLASSIFIERS}
\name{Mikhail Silaev$^{1}$, Konstantinos Drossos$^{2}$, and Tuomas Virtanen$^{1}$}
\address{
$^{1}$Tampere University, Tampere, Finland\\
$^{2}$Nokia Technologies, Espoo, Finland
}

\begin{document}
%\ninept

\maketitle

\begin{abstract}
Generative adversarial networks (GANs) and diffusion models have recently achieved state-of-the-art performance in audio super-resolution (ADSR), producing perceptually convincing wideband audio from narrowband inputs. However, existing evaluations primarily rely on signal-level or perceptual metrics, leaving open the question of how closely the distributions of synthetic super-resolved and real wideband audio match.
Here we address this problem by analyzing the separability of real and super-resolved audio in various embedding spaces. We consider both middle-band ($4\to 16$~kHz) and full-band ($16\to 48$~kHz) upsampling tasks for speech and music, training linear classifiers to distinguish real from synthetic samples based on multiple types of audio embeddings.
Comparisons with objective metrics and subjective listening tests reveal that embedding-based classifiers achieve near-perfect separation, even when the generated audio attains high perceptual quality and state-of-the-art metric scores. This behavior is consistent across datasets and models, including recent diffusion-based approaches, highlighting a persistent gap between perceptual quality and true distributional fidelity in ADSR models. 
 Code and demo are available at \url{https://github.com/msilaev/ADRS} .

\end{abstract}

\section{Introduction}
\label{sec:introduction}
% 0.8 pages

 The primary objective of generative adversarial networks (GANs)~\cite{goodfellow2014generative} is to generate synthetic ('fake') samples that closely resemble real data, effectively sampling from a distribution approximating the real one. 
 {Evaluating how well GANs achieve this goal remains a fundamental challenge, as traditional metrics often fail to capture perceptual quality, diversity and overfitting to the training data. Reliable evaluation is crucial for understanding model behavior and guiding improvements in generative quality. }
To evaluate GAN performance, various approaches have been proposed. For example, the representational quality of a GAN's discriminator is evaluated by reusing its frozen intermediate layers as feature extractors for supervised downstream tasks, such as image classification~\cite{radford2015unsupervised}. Classifiers trained on learned representation embeddings have been used to measure privacy preservation in the speech and audio domains, including automatic speech recognition~\cite{srivastava2019privacy} and sound event detection~\cite{gharib2024adversarial}.

 %%%%%%%%%%%%%%%%%%%%%%%%%%%%%%%%%%%%%%%%%%%%%
 \begin{figure}[t!]
 \centering
 \includegraphics[width=0.7\linewidth]{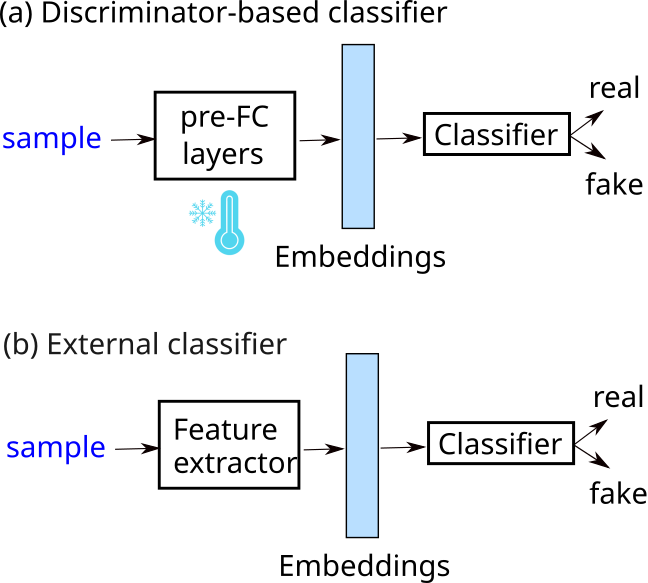} 
 \caption{
Feature extraction and 'real'/'fake' classification task. 
(a) Discriminator-based classifier uses internal discriminator features produced by its pre-fully connected (pre-FC) layers with frozen weights.
%It is different from the discriminator itself, as pre-FC layers are frozen and only the classifier layer is trained on the entire dataset. 
(b) External classifier operating on independent features extracted by a separate network, enabling an analysis of potential representation bias between the two classifiers. 
 }
 \label{Fig:EmbeddingsClassifiers}
\end{figure}

A complementary direction compares the statistical similarity of real and generated samples in embedding spaces such as that induced by a pretrained Inception network~\cite{heusel2017gans, lucic2018gans, sajjadi2018assessing}. Such embeddings enable quantitative metrics like the Fréchet Inception Distance~\cite{heusel2017gans}, Fréchet Audio Distance~\cite{kilgour2018fr}, Deep Speech Distance~\cite{binkowski2019high}, and precision–recall measures~\cite{lucic2018gans, sajjadi2018assessing, kynkaanniemi2019improved}, which assess the overlap and coverage between real and synthetic data manifolds.

However, these evaluation strategies have not been systematically applied to GAN-based super-resolution tasks, either in image~\cite{ledig2017photo} or audio~\cite{su2021bandwidth} domains. In this work, we extend the assessment of GANs and recently proposed diffusion models~\cite{im2025flashsr, yun2025flowhigh} by addressing a straightforward yet practically important question: Can linear classifiers trained on either frozen discriminator representations (Fig.~\ref{Fig:EmbeddingsClassifiers}a) or external embeddings (Fig.~\ref{Fig:EmbeddingsClassifiers}b) reliably distinguish real wideband from super-resolved audio samples?

We focus on the practically relevant audio super-resolution (ADSR) problem, which aims to enhance low-bandwidth audio signals by generating new high-frequency content, thereby expanding their spectral range. This topic has attracted considerable research interest in recent years~\cite{im2025flashsr,yun2025flowhigh,gupta:2022:spcom, liu:2024:icassp, li2019speech, moliner2022behm, kim2019bandwidth, su2021bandwidth, kong2020hifi, Li2021RealTimeBWE, kuleshov2017audio}.
The evaluation of GANs and diffusion models for ADSR has focused on signal-level metrics such as signal-to-noise ratio (SNR), log-spectral distance (LSD), and perceptual measures including mean opinion score (MOS) and subjective listener tests~\cite{su2021bandwidth, moliner2022behm}. In contrast, GAN-specific evaluations have not yet been explored. This gap is particularly relevant for full-band ADSR~\cite{su2021bandwidth}, which is the primary focus of the present work, where conventional objective metrics and human listeners often struggle to capture the subtle, yet perceptually important differences between real and synthesized signals. 

The rest of the paper is organized as follows. The GAN-based
ADSR and architectures are outlined in Sec.~\ref{Sec:SDSRusingGANs}. 
The concept of 'real'/'fake' classifiers is explained in Sec.~\ref{Sec:RepresentationBias}. 
Evaluation results including objective metrics,  
 MUSHRA (Multi-Stimulus Test with Hidden Reference
and Anchor) listening test \cite{ITU-R-BS1534-3} for the 
full-band ADSR and classifier accuracy are reported in Sec.~\ref{Sec:Evaluation}. Conclusions are presented in Sec.~\ref{Sec:Conclusions}.

\section{Audio Super-Resolution Models}
\label{Sec:SDSRusingGANs}

Given audio signal $\bm{x} = [x(n/f_s)]$, $n=0,...,N$ the ADSR task aims to reconstruct an upsampled signal $\bm{y} = [y(m/r f_s)]$, $m=0,...,rN$, where $r$ is the upsampling ratio. The input and upsampled signals thus have lengths $N$ and $rN$, respectively, but share the same duration $N/f_s$.
Increasing the sampling rate $f_s \rightarrow r f_s$ expands the Nyquist frequency from $f_s/2$ to $r f_s/2$, allowing new frequency components in the band $[f_s/2, r f_s/2]$. The original and upsampled signals are referred to as narrowband (NB) and wideband (WB), respectively.

Deep learning methods for ADSR have advanced rapidly, evolving from supervised models such as AudioUNet~\cite{kuleshov2017audio} to GAN-based and diffusion-based generative models~\cite{moliner2022behm, kim2019bandwidth, kong2020hifi, Li2021RealTimeBWE, hifi-gan-bwe}. 
In this work, we adopt the MU-GAN (Multiscale U-Net GAN) architecture, originally proposed for the 
$4\to16$ kHz~ADSR~\cite{kim2019bandwidth}. 
We further extend MU-GAN to the full-band $16\to48$~kHz ADSR.
Due to limited implementation details and the absence of publicly available code, we re-implemented, trained, and evaluated the model from scratch.
As a supervised baseline, we employ the AudioUNet model~\cite{kuleshov2017audio} for both $4\to16$~kHz and $16\to48$~kHz upsamplings. 
The trained MU-GAN discriminator features are used as embeddings for the 'real'/'fake' classification task across all considered models.

Both AudioUNet and MU-GAN models were implemented in PyTorch\footnote{\url{https://github.com/msilaev/ADRS}} and trained using the VCTK speech dataset~\cite{yamagishi2019cstr} for $4 \rightarrow 16$ kHz and $16 \rightarrow 48$ kHz ADSR, and the FMA-small music dataset~\cite{defferrard2017fma} for $16 \rightarrow 48$ kHz ADSR. 
{We used the official train/validation/test splits for the FMA dataset~\cite{defferrard2017fma} and the train/test split for the VCTK dataset, following the protocol used in previous works~\cite{su2021bandwidth,kim2019bandwidth,kuleshov2017audio}.}
Training was performed on a single NVIDIA A100 GPU. For the VCTK dataset, both models were trained for $500$ epochs. For the FMA-small dataset, AudioUNet and MU-GAN were trained for $100$ and $80$ epochs, respectively, with early stopping. The learning rate was set to $10^{-4}$ and the mini-batch size to 128. The generator and discriminator were optimized using Adam and stochastic gradient descent (SGD), respectively.
To stabilize adversarial training, we adopted a scheduled update strategy, where the generator was updated more frequently than the discriminator. 
For the converged MU-GAN models, the discriminator achieved an accuracy of approximately $51\%$ on the VCTK dataset and $49\%$ on the FMA dataset, indicating that the model reached the desired equilibrium characteristic of well-trained GANs, where real and generated samples become nearly indistinguishable.

To enable comparison with state-of-the-art methods for full-band ADSR, we use HiFi-GAN, originally proposed for denoising and bandwidth extension~\cite{su2021bandwidth}, and two recent diffusion-based models, FlowHigh~\cite{im2025flashsr} and FlashSR~\cite{yun2025flowhigh}, both capable of upscaling audio from arbitrary input sampling rates to $48$~kHz. For our experiments, we use the publicly available unofficial HiFi-GAN implementation with pretrained weights for inference~\cite{hifi-gan-bwe}, and the official inference code released for FlowHigh\footnote{\url{https://github.com/resemble-ai/flowhigh}}
 and FlashSR\footnote{\url{https://github.com/jakeoneijk/FlashSR_Inference}}.

%%%%%%%%%%%%%%%%%%%%%%%%%%%%%%%%%%

\section{'Real' / 'Fake' Audio Classifiers}
\label{Sec:RepresentationBias}

Classifiers capable of distinguishing real and synthetic audio clips are trained and evaluated using several labeled embedding datasets.
These datasets are constructed by transforming real and synthetic audio signals into fixed-length embeddings using different feature extractors. Several types of embeddings are considered. 
First, to study the quality of learned representations, we use features extracted from the fixed pre-FC layer of the MU-GAN discriminator (Fig.~\ref{Fig:EmbeddingsClassifiers}a). In our implementation, an $8192$-sample input is mapped to a $32$-dimensional embedding vector by this discriminator layer.

In addition, two types of external embeddings are employed (Fig.~\ref{Fig:EmbeddingsClassifiers}b). The OpenL3~\cite{cramer2019look} model generates a $512$-dimensional embedding vector from a $1$-second audio segment. This model is suitable for the $4 \to 16$~kHz ADSR task but cannot be reliably applied to the $16 \to 48$~kHz setting due to its limited input bandwidth, which prevents it from capturing the full frequency content of $48$~kHz audio.

To address the bandwidth limitations of OpenL3, log-Mel spectrogram energies are used as an alternative feature representation.
We use $256$ Mel-frequency bins, an FFT size of $4096$, and a hop length of $256$.
The upper frequency limit is set by the Nyquist frequency, corresponding to $8$~kHz and $24$~kHz for input sampling rates of $16$~kHz and $48$~kHz, respectively.
To produce fixed-length embeddings that are independent of input duration, adaptive average pooling is applied along the temporal dimension.

\begin{table}[htb!]
    \caption{ LSD and SNR metrics for different models  }
    \label{tab:metrics_combined}
    \centering
    \begin{tabular}{l@{\hskip 4pt}c@{\hskip 4pt}c@{\hskip 4pt}c@{\hskip 4pt}c@{\hskip 4pt}c@{\hskip 4pt}c}
        \toprule
        & \multicolumn{4}{c}{\textbf{VCTK}} & \multicolumn{2}{c}{\textbf{FMA}} \\
        \cmidrule(lr){2-5} \cmidrule(lr){6-7}
        \textbf{Model} & \textbf{LSD} & \textbf{SNR} & \textbf{LSD} & \textbf{SNR} & \textbf{LSD} & \textbf{SNR} \\
        & 4$\to$16 & 4$\to$16 & 16$\to$48 & 16$\to$48 & 16$\to$48 & 16$\to$48 \\
        \midrule
        %Spline      & 9.9  & 13.6 & 10 & 21  & --   & --    \\
        AudioUnet   & 4.5  & 15.4 & 4.2  & 22  & 9.2  & 24.5  \\
        MU-GAN      & 3.9  & 14.6 & 4.2  & 20.8  & 6.7  & 27.3  \\
        HiFi-GAN    & --   & --   & 2.1  & 17.5  & --   & --    \\
        FlowHigh    & --   & --   & 3  & -6.8  & 3.6  & -3  \\
        FlashSR     & --   & --   & 3.9  & 16  & 8.4  & 18.2  \\
        \bottomrule
    \end{tabular}
\end{table}

%%%%%%%%%%%%%%%%%%%%%%%%%%%%%%%%%%%%%%
\section{Evaluation and results}
\label{Sec:Evaluation}

To evaluate the performance and perceived realism of ADSR models, we combine objective signal comparison metrics, signal-to-noise ratio (SNR), and logarithmic spectral distance (LSD) with MUSHRA listening tests.

%%%%%%%%%%%%%%%%%%%%%%%%%%%%%%
\subsection{Evaluation of Signal-Level Performance} 
Table~\ref{tab:metrics_combined} summarizes the SNR and LSD results, mainly to show that no unexpectedly low or high values are observed. The $4\to16$~kHz ADSR results are consistent with previous work~\cite{kim2019bandwidth,kuleshov2017audio}, and, as in those studies, LSD is computed using the natural logarithm.

A notable observation is that the FlowHigh model yields negative SNR values while achieving lower LSD than AudioUNet and MU-GAN. The negative SNR does not indicate degraded perceptual quality but instead arises from a global amplitude scaling factor, which may vary across samples.

No significant differences are observed between HiFi-GAN, AudioUNet, and MU-GAN scores. However, as shown below, the SNR and LSD metrics correlate poorly with human listening results, which reveal clear perceptual differences between models' outputs.

%%%%%%%%%%%%%%%%%%%%%%%%%%%%%%%%%%%%%%%%%%%%
\begin{figure}[htb!]
  \centering
  \begin{tabular}{cc}
  \put(-60,0){    (a) } 
  \put(60,0){  (b)  }
  \\
  \includegraphics[width=1.0\linewidth]{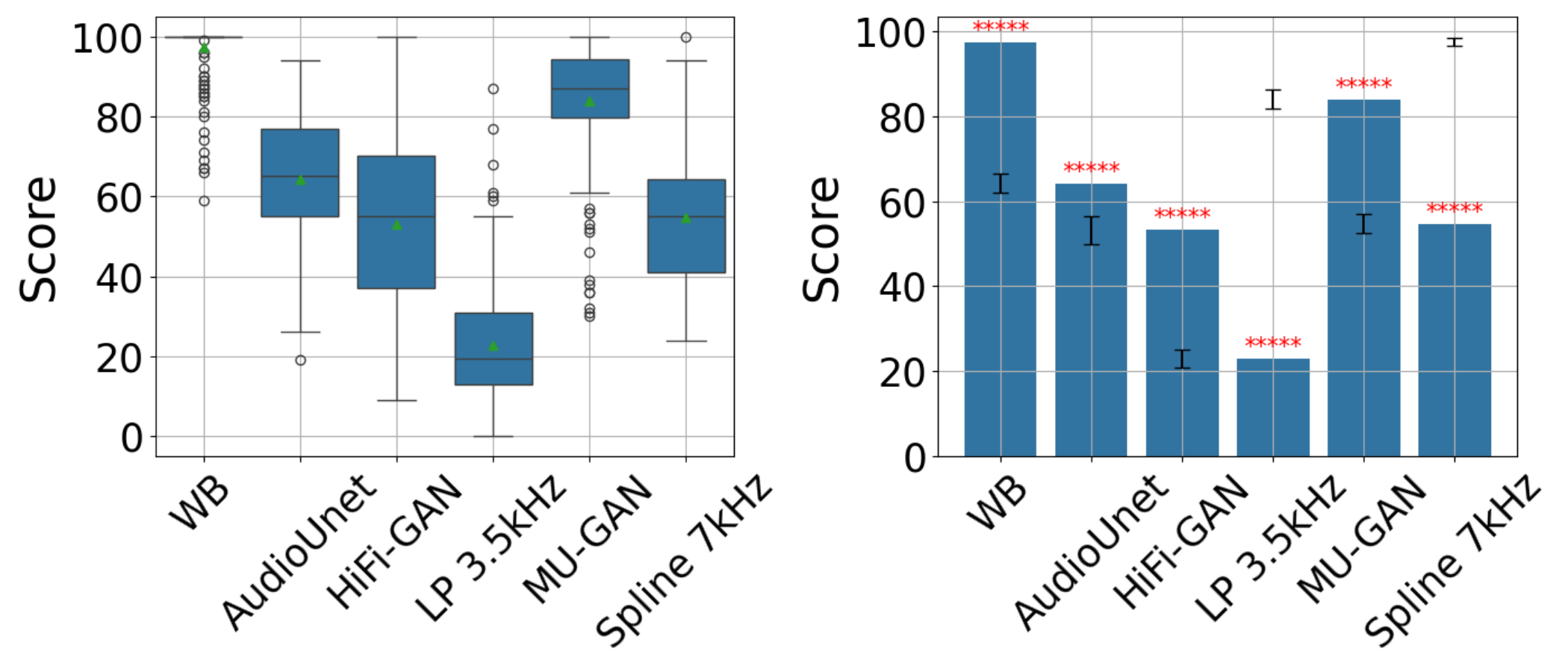}
  \end{tabular} %\\
  \caption[MUSHRA scores for ADSR methods]{
     Listener scores for different ADSR methods.
    MUSHRA scores for listening tests across different conditions WB, MU-GAN, AudioUnet, HiFi-GAN, LP 3.5kHz, Spline-Up 7kHz.
    (a)  Inter-quartile range (IQR), medians, and mean values by green triangles. 
    (b) Mean scores with error bars representing $95\%$
    confidence intervals.
    }   
    \label{Fig:MUSHRA}
\end{figure}

%%%%%%%%%%%%%%%%%%%%%%%%%%%%%%%%%%%%%%%%%

\subsection{MUSHRA listening test }
\label{subsec:mushra}
For evaluation, $12$ recordings were randomly selected from the VCTK test set, ensuring an equal number of male and female speakers.
Original WB recordings at $48$ kHz were down-sampled to $16$ kHz and then up-sampled back to $48$ kHz using three different ADSR models: AudioUnet, MU-GAN, and HiFi-GAN.
Each recording was supplemented by two anchor signals: a low-pass filtered version at $3.5$ kHz and a middle-pass filtered version at $7$ kHz. Furthermore, the original NB recording was included. This yields six experimental conditions.
The listening test was implemented using the MUSHRA listening test interface, which allowed listeners to set loops if they wanted to focus on particular short passages of the audio signal.

Listening test results are shown in Fig.~\ref{Fig:MUSHRA} following the MUSHRA recommendations~\cite{ITU-R-BS1534-3}. The box plot in Fig.~\ref{Fig:MUSHRA}a shows the distribution of MUSHRA scores for six conditions: WB, MU-GAN, AudioUnet, HiFi-GAN, LP 3.5 kHz, and Spline-Up $7$ kHz, highlighting median, IQR, mean values, and outliers. The bar graph in Fig.~\ref{Fig:MUSHRA}b shows the mean scores with $95\%$ confidence intervals. MU-GAN achieves the highest score, closely matching the WB reference, while HiFi-GAN performs the worst, closely resembling the $7$ kHz anchor. AudioUnet performs slightly better. 
%These results show that MU-GAN outperforms traditional upsampling and deep learning approaches for full-band ADSR tasks.
Non-overlapping confidence intervals in Fig.~\ref{Fig:MUSHRA}b show that listeners can reliably distinguish between real and synthetic audio, even though the perceptual quality of the MU-GAN outputs remains comparable to that of the target WB recordings. 
In the next section, we demonstrate that this distinction is also captured by binary classifiers.

\subsection{‘Real’/‘Fake’ Classifier Accuracy}

The embedded datasets are constructed using audio clips from the test subsets of the VCTK (speech) and FMA-small (music) datasets, which were not seen during the training of any considered model. Each embedding dataset is randomly shuffled and split into training ($80\%$) and testing ($20\%$) subsets.
Before classification, the feature vectors are standardized to have zero mean and unit variance.
A linear discriminant analysis (LDA) classifier is then trained on the normalized training data and evaluated on the test data to assess its ability to distinguish between real and fake samples.
Additionally, the test embeddings are projected into the discriminant subspace learned by LDA, where class separation is maximized.
This projection enables a qualitative visualization of class overlap and separability along a single discriminant axis.

\begin{figure}[!htb]
\begin{minipage}[t]{.49\linewidth}
  \centering
  \includegraphics[width=4.0cm]{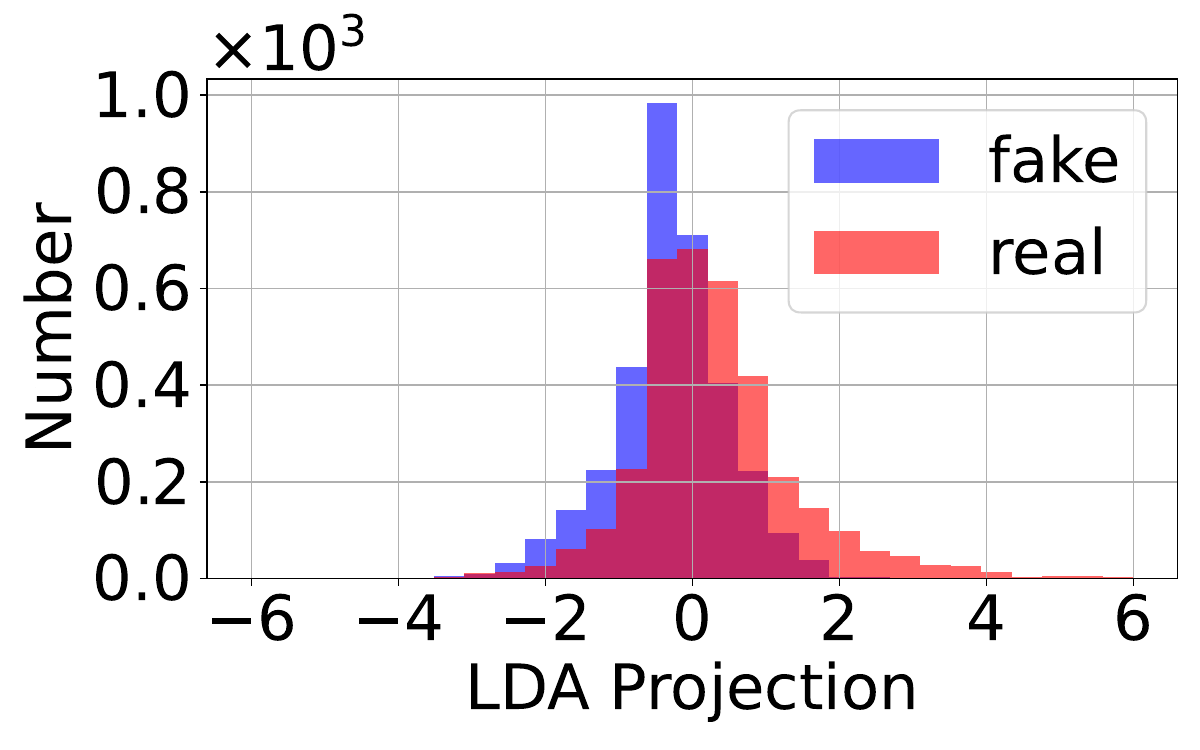}
  \subcaption{\footnotesize MU-GAN 16 kHz, Discriminator }
  \label{fig:lda_gan_16_disc}
\end{minipage}
\hfill
\begin{minipage}[t]{.49\linewidth}
  \centering
  \includegraphics[width=4.0cm]{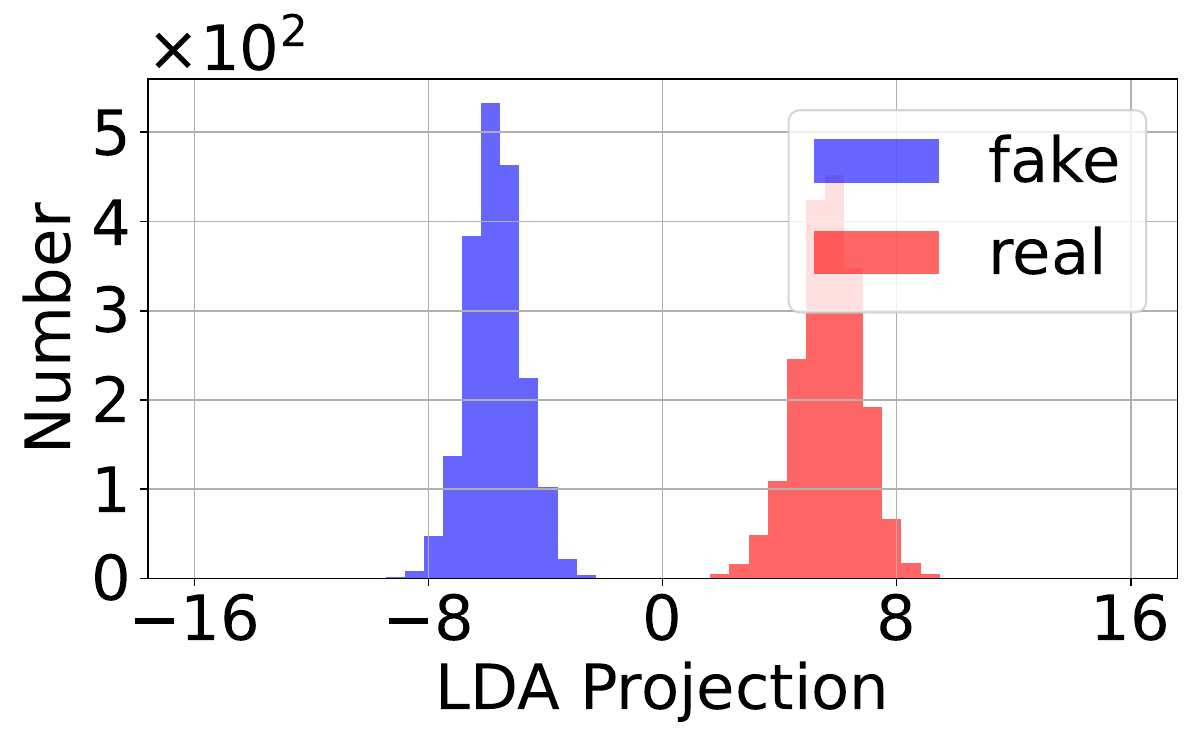}
  \subcaption{\footnotesize MU-GAN 16 kHz, OpenL3 }
  \label{fig:lda_gan_16_openl3}
\end{minipage}
\vspace{0.1cm}
%%%%%%%%%%%%%%%%%%%5
\begin{minipage}[t]{.49\linewidth}
  \centering
  \includegraphics[width=4.0cm]{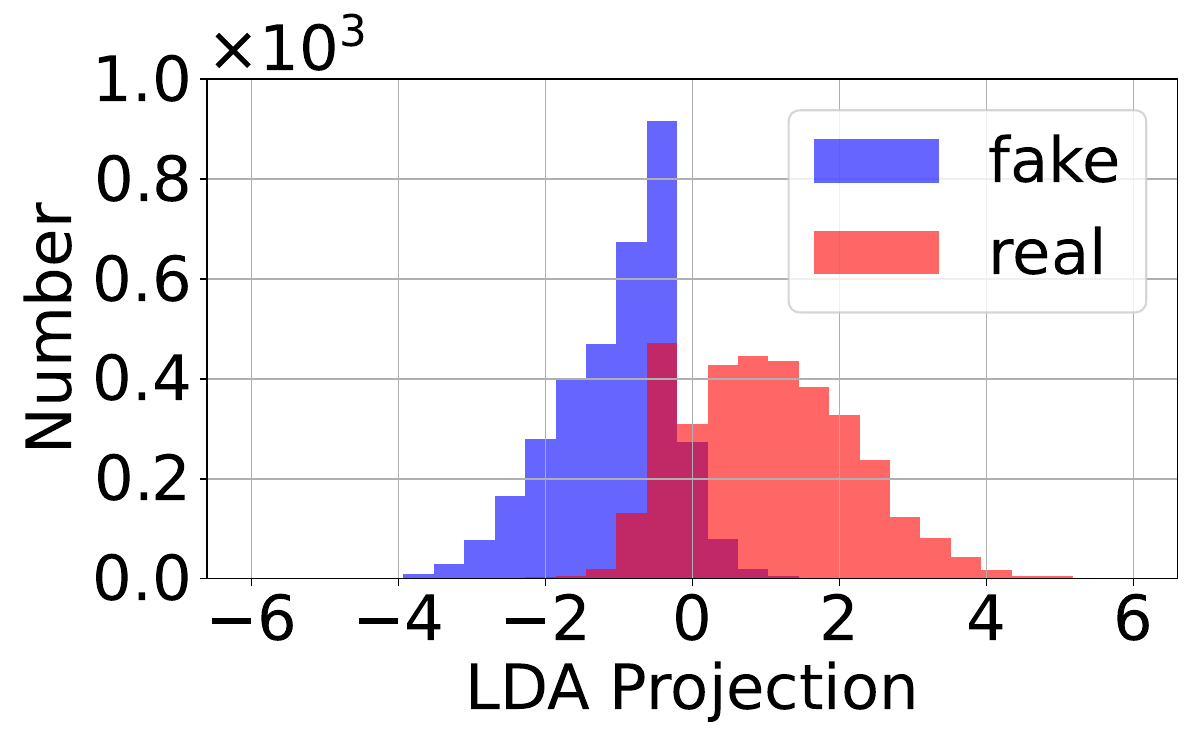}
  \subcaption{\footnotesize AudioUnet 16 kHz, Discriminator }
  \label{fig:lda_audiounet_16_disc}
\end{minipage}
\hfill
\begin{minipage}[t]{.49\linewidth}
  \centering
  \includegraphics[width=4.0cm]{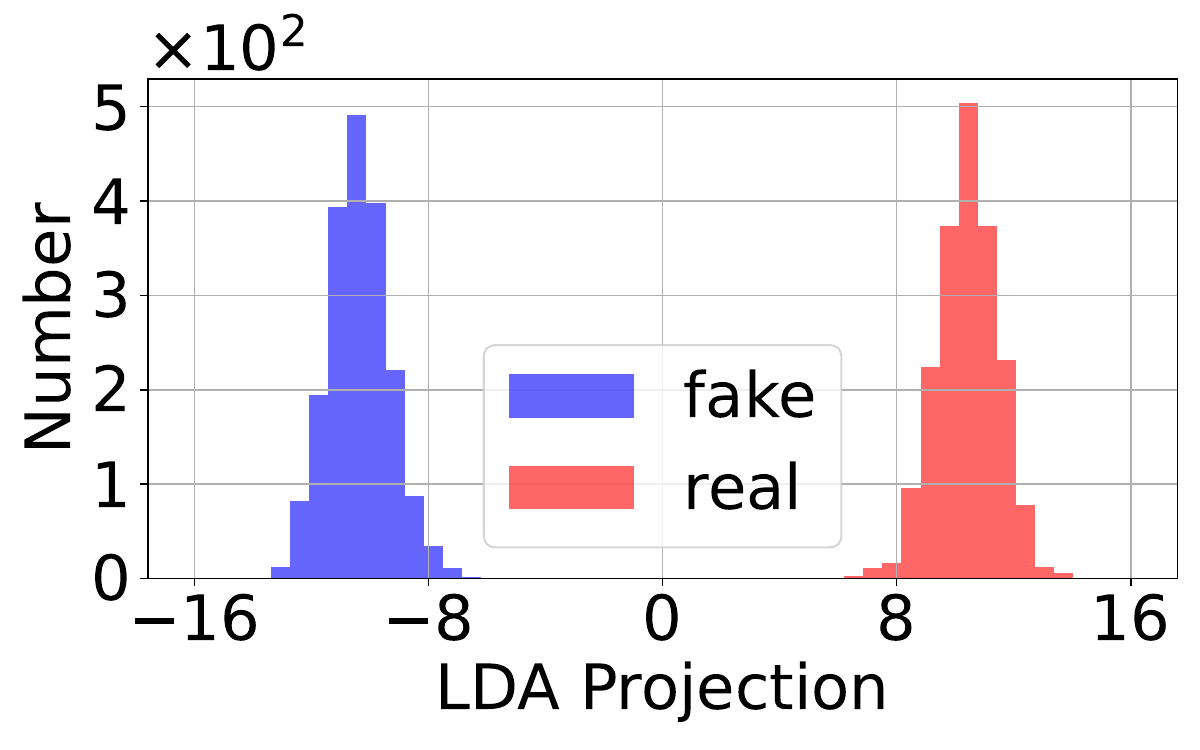}
  \subcaption{\footnotesize AudioUnet 16 kHz, OpenL3 }
  \label{fig:audiounet_16_openl3}
\end{minipage}
\vspace{0.1cm}
%%%%%%%%%%%%%%%%%%%5
\begin{minipage}[t]{.49\linewidth}
  \centering
  \includegraphics[width=4.0cm]{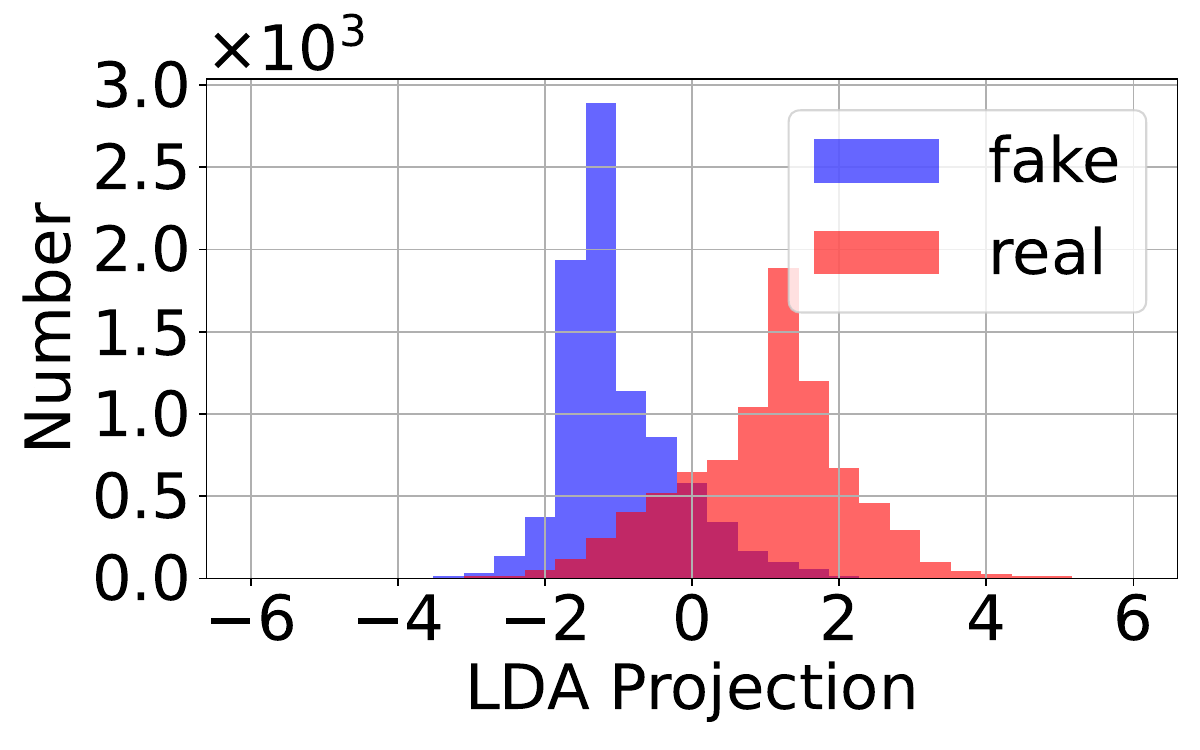}
  \subcaption{ \footnotesize MU-GAN 48 kHz, Discriminator}\label{fig:lda_gan_disc}
\end{minipage}
\hfill
\begin{minipage}[t]{.49\linewidth}
  \centering
  \includegraphics[width=4.0cm]{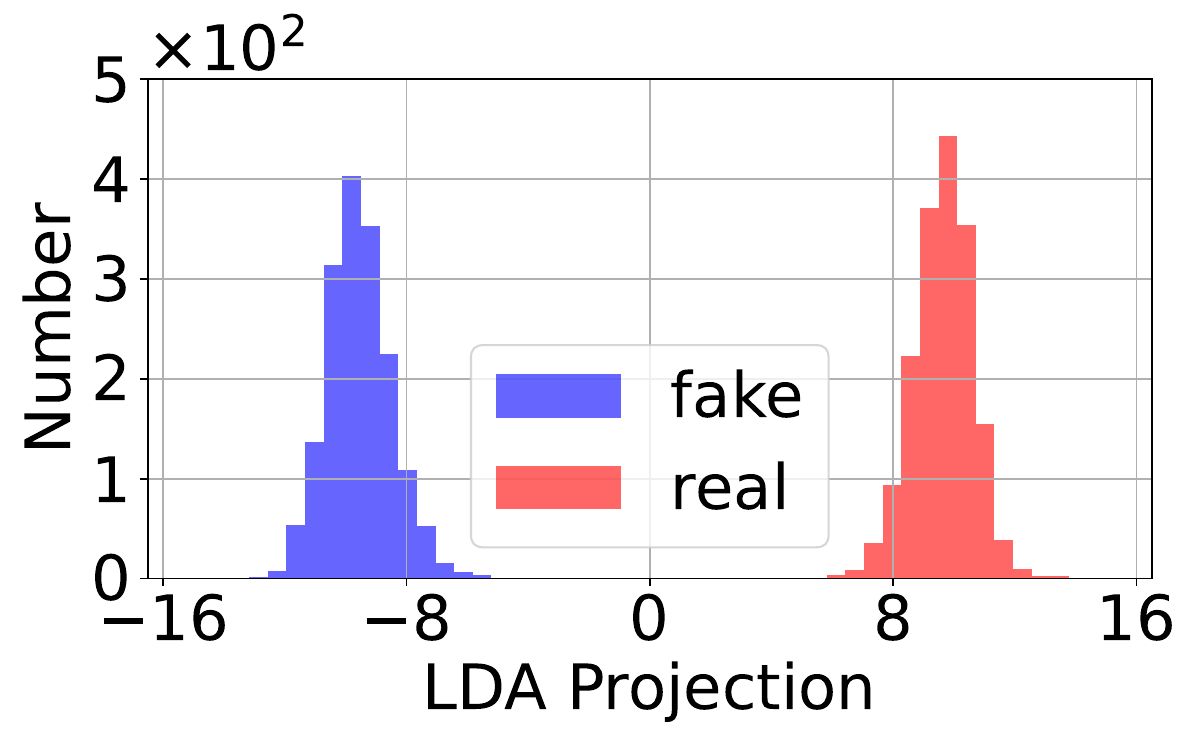}
  \subcaption{ \footnotesize MU-GAN 48 kHz, log-Mel}
  \label{fig:lda_gan_mel}
\end{minipage}
\vspace{0.1cm}
\begin{minipage}[t]{.49\linewidth}
  \centering
  \includegraphics[width=4.0cm]{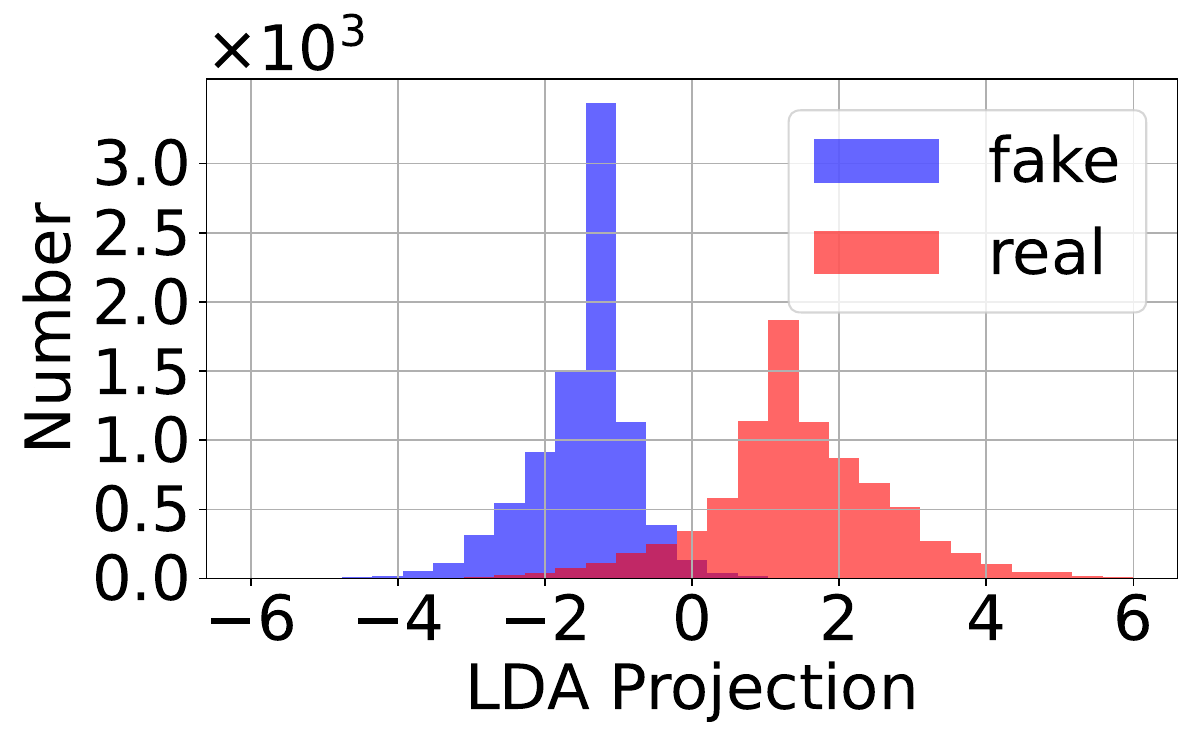}
  \subcaption{ \footnotesize AudioUnet 48~kHz, Discriminator}\label{fig:lda_audiounet_disc}
\end{minipage}
\hfill
\begin{minipage}[t]{.49\linewidth}
  %\centering
  \includegraphics[width=4.0cm]{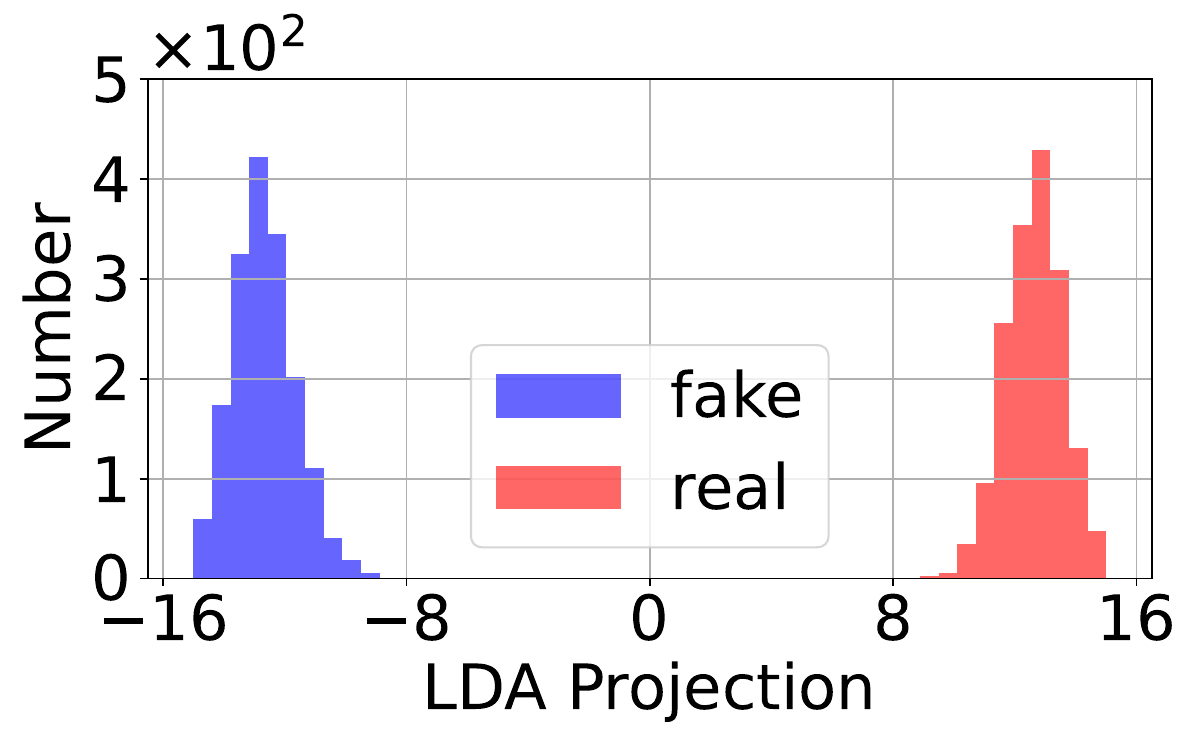}
  \subcaption{ \footnotesize AudioUnet 48~kHz, log-Mel}
  \label{fig:lda_audiounet_mel}
\end{minipage}
\caption{LDA projections for audio clips from VCTK test dataset, calculated using  MU-GAN discriminator features (left column), OpenL3 (right column b,d) and log-Mel (right column f,h) embeddings. (a-d) $4 \to 16$~kHz  and (e-h) $16\to48$~kHz ADSR. 
}
\label{Fig:lda_48}
\end{figure}

Example LDA distributions for different embedding spaces are shown in Fig.~\ref{Fig:lda_48}.
Panels corresponding to (a,c,e) discriminator pre-FC feature embeddings show completely overlapping distributions for the $4\to16$~kHz MU-GAN, partial overlap for the $16\to48$~kHz MU-GAN, and almost complete separation for AudioUNet.
On the one hand, this demonstrates that the discriminator learns useful features capable of distinguishing real from synthetic clips.
On the other hand, the clear difference in distribution overlap between MU-GAN and AudioUNet indicates that the generator is trained to make these distributions more similar. 

In contrast, panels (b,d) and (f,h) show that LDA projections of the OpenL3 and log-Mel embeddings demonstrate complete 'real'/'fake' class separation, achieving $100\%$ classification accuracy. Note that panels (e,f) and (g,h) correspond to MU-GAN and AudioUnet models for $16\to48$~kHz upsampling studied using MUSHRA test reported in Sec.~\ref{subsec:mushra}.

\begin{table}[!htb]
    \caption{
    Binary 'fake'/'real' classifier accuracy (\%) based on MU-GAN discriminator embeddings. Columns correspond to ADSR  $4\to 16$~kHz on VCTK, $16\to 48$~kHz on VCTK, $16\to 48$~kHz on FMA-small datasets. }
    \label{tab:LDAaccuracy_combined}
    \centering
    \setlength{\tabcolsep}{3pt} % reduce column spacing
    \begin{tabular}{lccc}
        \toprule
        \textbf{Model} & \textbf{VCTK 4$\to$16} & \textbf{VCTK 16$\to$48} & \textbf{FMA 16$\to$48} \\
        \midrule
        AudioUnet  & 80\% & 95\% & 78\% \\
        MU-GAN     & 56\% & 83\% & 70\% \\
        HiFi-GAN   & --   & 93\% & --   \\
        FlowHigh   & --   & 85\% & 74\% \\
        FlashSR    & --   & 88\% & 66\% \\
        \bottomrule
    \end{tabular}
\end{table}

These results hold qualitatively for all models considered.
As shown in Table~\ref{tab:LDAaccuracy_combined}, the learned MU-GAN discriminator embeddings yield high accuracies (around $90\%$) for the AudioUNet, FlashSR, and FlowHigh models on the VCTK dataset, and slightly lower (around $70\%$–$80\%$) for the FMA dataset.
At the same time, the log-Mel and OpenL3 embeddings achieve perfect classification accuracy ($100\%$) across all tasks and models considered.

%%%%%%%%%%%%%%%%%%%%%%%%%%%%%%%%%%%%%%%%%%%%%%
\section{Conclusions}
\label{Sec:Conclusions}

This study highlights a gap between traditional signal metrics, perceptual quality, and the separability of real and super-resolved audio distributions produced by generative ADSR models. During stable GAN training, the discriminator accuracy converges to around $50\%$, yet its learned representations can still distinguish 'real' and 'fake' audio in a supervised benchmark, indicating that the GAN captures comprehensive data features. Listener scores in the MUSHRA test closely match the target wideband audio.

Classifiers trained on external embeddings, such as OpenL3 or log-Mel spectrograms, achieve nearly perfect separation  between real and generated clips. This behavior is consistent across domains, sampling rates ($4\to16$~kHz and $16\to48$~kHz), and extends to state-of-the-art diffusion models~\cite{im2025flashsr,yun2025flowhigh}. These results suggest that high perceptual quality does not necessarily imply accurate distribution modeling, revealing a persistent gap between perceptual realism and representational fidelity—an open challenge for future ADSR research.

\bibliographystyle{IEEEbib}
\bibliography{refs_2}

\end{document}